\documentclass[showpacs,aps,prc,reprint,nofootinbib]{revtex4-1}
       \usepackage{color}
\usepackage{epsfig}
\usepackage{graphicx}
\usepackage{amssymb}
\begin{document}
\title{Revisiting an extended-mean-field approach in heavy-ion collisions around the Fermi energy}
\author{G.\ Besse$^1$}
\author{V.\ de la Mota$^1$}
\email{delamota@subatech.in2p3.fr}
\author{E.\ Bonnet$^1$}
\author{P.\ Eudes$^1$}
\email{eudes@subatech.in2p3.fr}
\author{P.\ Napolitani$^2$}
\author{Z.\ Basrak$^3$}
\affiliation{$^1$SUBATECH, Universit\'e de Nantes/IMT-Atlantique/CNRS,
          F-44\,307 Nantes Cedex, France}
\affiliation{$^2$Universit\'e Paris-Saclay, CNRS/IN2P3, IJCLab,
          91\,405 Orsay, France}
\affiliation{$^3$Ru{d\llap{\raise 1.22ex\hbox
   {\vrule height 0.09ex width 0.2em}}\rlap{\raise 1.22ex\hbox
   {\vrule height 0.09ex width 0.06em}}}er
   Bo\v{s}kovi\'{c} Institute, HR-10\,002 Zagreb P.P.\ 180, Croatia}
\date{\today}
\begin{abstract}
Static and dynamical aspects of nuclear systems are described through an extended-time dependent mean-field approach. The foundations of the formalism are presented, with highlights on the estimation of average values and their corresponding dispersions. In contrast to semiclassical transport models the particular interest of this description lies on its intrinsic quantal character. The reliability of this approach is discussed by means of stopping-sensitive observables analysis in heavy ion collisions in the range of 20 to 120 MeV per nucleon.
\end{abstract}
\maketitle
\section{\label{sec1}Introduction}
One of the fundamental microscopic models in nuclear and atomic physics is the time-dependent Hartree-Fock (TDHF) approximation. 
since the pioneer works in the nuclear field, it has been largely applied to the description of collective phenomena in heavy ion collisions at incident energies below 10 MeV per nucleon \cite{tdhf}. 
Beyond this energy regime this approximation turned out to be insufficient to correctly describe the dissipative processes observed in experiments. 
For this reason, quantal and dynamical extensions have been developed in order to account for those dynamical correlations which are lacking in mean field approaches. 
In the range of energies we are interested in, the so-called extended-time-dependent Hartree-Fock (ETDHF) approach, either quantal or semiclassical, considers the residual interactions by the inclusion of a collision term: The mean field evolution equation for the one-body density matrix is complemented by a Boltzmann (or Uehling-Uhlenbeck) term\ \cite{etdhf}. 
These models have been shown to successfully describe the irreversible mean behavior of nuclear systems towards equilibrium. 
Nevertheless, with the improvement of the quality and completeness of experimental data these extended theories attained their limits too when they tried to describe, on one side, the variety of channels observed experimentally and, on the other, the dispersion of observables.
The essential problem of ETDHF theories was the lack of density fluctuations, large  enough to account for those phase space bifurcations giving rise to the observed manifold of  exit channels.
Also the absence of small density fluctuations needed to draw the statistical dispersions on one-body observables is a drawback of these kind of descriptions.
The search of a convenient description of those aspects has been a challenge for nuclear many-body theories for a long time, and much work has been done in order to modelize the effects of many-body correlations in extended-mean-field theories, namely by including a fluctuation force of the Langenvin type \cite {bl}.
The task of including a fluctuating term in our ETDHF description is then mandatory and it is foreseen for further publications.
Nevertheless, before undertaking this work it is imperative to provide an overview of the state of the art of the model (the most recent developments, reviewing the latest ideas and features), which constitutes the starting point of next developments.

This work is organized as follows: in Sec.\ \ref{sec2} the theoretical formalism is presented: Subsection A is devoted to the characteristics of the initial state, B and C to the fundamental aspects of mean-field and dissipative dynamics, respectively. In subsection D a scheme for the treatment of observable dispersions is displayed. In Sec.\ \ref{sec3} we present the results concerning stopping power observables measuring the degree of dissipation for different systems. Finally the conclusions and perspectives are presented in Sec.\ \ref{sec4}.

\section{\label{sec2}The ETDHF formalism}
In this section the bases of the model are revisited.
The first version of the model, which was initially called dynamical wavelets in nuclei (DYWAN) dates back to 1998 \cite{dywan}. 
In that paper the theoretical background was presented starting from the derivation of the quantal kinetic Boltzmann-like equation of motion for the one-body density matrix by means of projection methods.
Due to its complexity, solving this equation is, however, a difficult task. 
Thus, the search for a convenient resolution scheme is needful in order to establish an efficient and reliable treatment of the evolution of states and relevant observables. 
The wavelet theory has been considered with the aim of obtaining an efficient and adaptive representation of the physical space corresponding to the many body system. 
In that work a biorthogonal spline basis was employed to span single-particle (SP) wave functions in a harmonic oscillator well. 
A one-to-one correspondence between the level of the description (pure mean-field, extended-mean-field and beyond) and the level in a multiresolution analysis is established and an iterative procedure is performed in order to optimize the number of coefficients in the SP wave functions' expansion. 
This procedure permits one to fix the optimum number, location, and widths of wavelets in phase space. 
It has been shown mathematically \cite{spline} that the spline wavelets can be expressed as linear combinations of squeezed coherent states (SCSs). 
The authors considered that property with particular interest since it brings the opportunity to make a more direct link with current dynamical models of heavy-ion collisions which use Gaussian states as projection elements. 
A set of spline wavelets were then implemented with the aim of performing an analysis of wave functions and the subsequent decomposition in SCSs [or simply coherent states (CSs)]. 
Coordinate representation of a three dimensional CS is
\begin{eqnarray}
\alpha(\vec r)&=&\alpha_x(x)\alpha_y(y)\alpha_z(z) , \\
\alpha_x(x)&=&\mathcal{N} \mathrm{exp}\lbrace - \lambda(x-\langle x \rangle)^2+i\langle k_x \rangle (x-\langle x \rangle) \rbrace , 
\end{eqnarray}
with similar expressions for $\alpha_y(y)$ and $\alpha_z(z)$. Here $\mathcal{N}=\left( \frac{1}{2\pi\chi} \right)^{1/4}$, the quantity $\lambda$ is defined as: $\lambda=\frac{\hbar}{2\chi} \left(1-i\frac{2\sigma}{\hbar}\right)$,
$\chi=\langle (x-\langle x \rangle)^2\rangle$ is the spatial second moment and $\sigma=\langle (x-\langle x \rangle)(k_x-\langle k_x \rangle)\rangle$ is the momentum-space correlation. 
SP wave functions $\varphi_{\lambda}$ and the one-body density matrix $\rho $ then result in
\begin{eqnarray}
|\varphi_{\lambda} \rangle(t) & = & \sum_{i=1}^{m_{\lambda}} c_i^{\lambda} |\alpha_i^{\lambda}\rangle(t) \label{varfi} , \\
\rho & = & \sum_{\lambda=0}^N \eta_{\lambda}|\varphi_{\lambda}\rangle \langle \varphi_{\lambda} | \simeq \sum_{\lambda=0}^N \sum_{i=1}^{m_{\lambda}} n_i^{\lambda} |\alpha_i^{\lambda}\rangle \langle \alpha_i^{\lambda}| , \label{ro1}
\end{eqnarray}
where $N$ is the number of nucleons, $m_{\lambda}$ is the number of CSs for a given SP state and $c_i^{\lambda}$ are constant coefficients, fixed at the initial time as $c_i^{\lambda}=1/\sqrt{m_{\lambda}}$, satisfying 
$$n_i^{\lambda}=|\sqrt{\eta_{\lambda}} c_i^{\lambda}|^2.$$
In Ref.\ \cite{dywan} nondiagonal matrix elements in $\rho$ are shown to be smaller and rapidly varying compared to diagonal ones and, with a good approximation, they can be neglected.
Since we deal with a three-dimensional problem, both the superscript $\lambda$ and the subscript $i$ in Eqs.\ (\ref{varfi}) and (\ref{ro1}) are in fact sets of three numbers. Indeed,
\begin{eqnarray}
|\varphi_{\lambda} \rangle(t) = |\varphi_{n_x}\varphi_{n_y}\varphi_{n_z}\rangle(t)
\end{eqnarray}
\noindent
then $\lambda=\lbrace n_x,n_y,n_z\rbrace$ represents the nucleon level or principal nucleon quantum number. The CS expansion of these functions gives
\begin{eqnarray}
|\varphi_{\lambda} \rangle= \sum_{i_x}\sum_{i_y}\sum_{i_z}  c_{i_x}^{n_x} c_{i_y}^{n_y}c_{i_z}^{n_z}  
|\alpha_{i_x}^{n_x}\alpha_{i_y}^{n_y}\alpha_{i_z}^{n_z}\rangle
\end{eqnarray}
which can be put in the form of Eq.\ (\ref{varfi}) with the following definitions: $i=\lbrace i_x,i_y,i_z\rbrace $,
$c_i^{\lambda}= c_{i_x}^{n_x} c_{i_y}^{n_y}c_{i_z}^{n_z}$, and $\alpha_i^{\lambda}=\alpha_{i_x}^{n_x}\alpha_{i_y}^{n_y}\alpha_{i_z}^{n_z}$.

In the framework of heavy-ion collisions around the Fermi energy, the dynamical evolution of the system is provided by the ETDHF equation
\begin{equation}\label{etdhf}
i\hbar\dot{\rho}= \ [ ~ \bf{h}, \rho \ ] + {\bf I}(\rho)
\end{equation}
\noindent
which presupposes that mean-field and residual interaction timescales are well separated. The former is assumed to be a slowly varying function of time compared to individual collisions rates.  Accordingly, the mean-field evolution and the collision integral are both computed in a self-consistent procedure but treated in a different manner from the numerical point of view. These aspects, which constitute the foundations of the model, are revisited in the following subsections.

\begin{figure}[tb]
\includegraphics[width=8.cm]{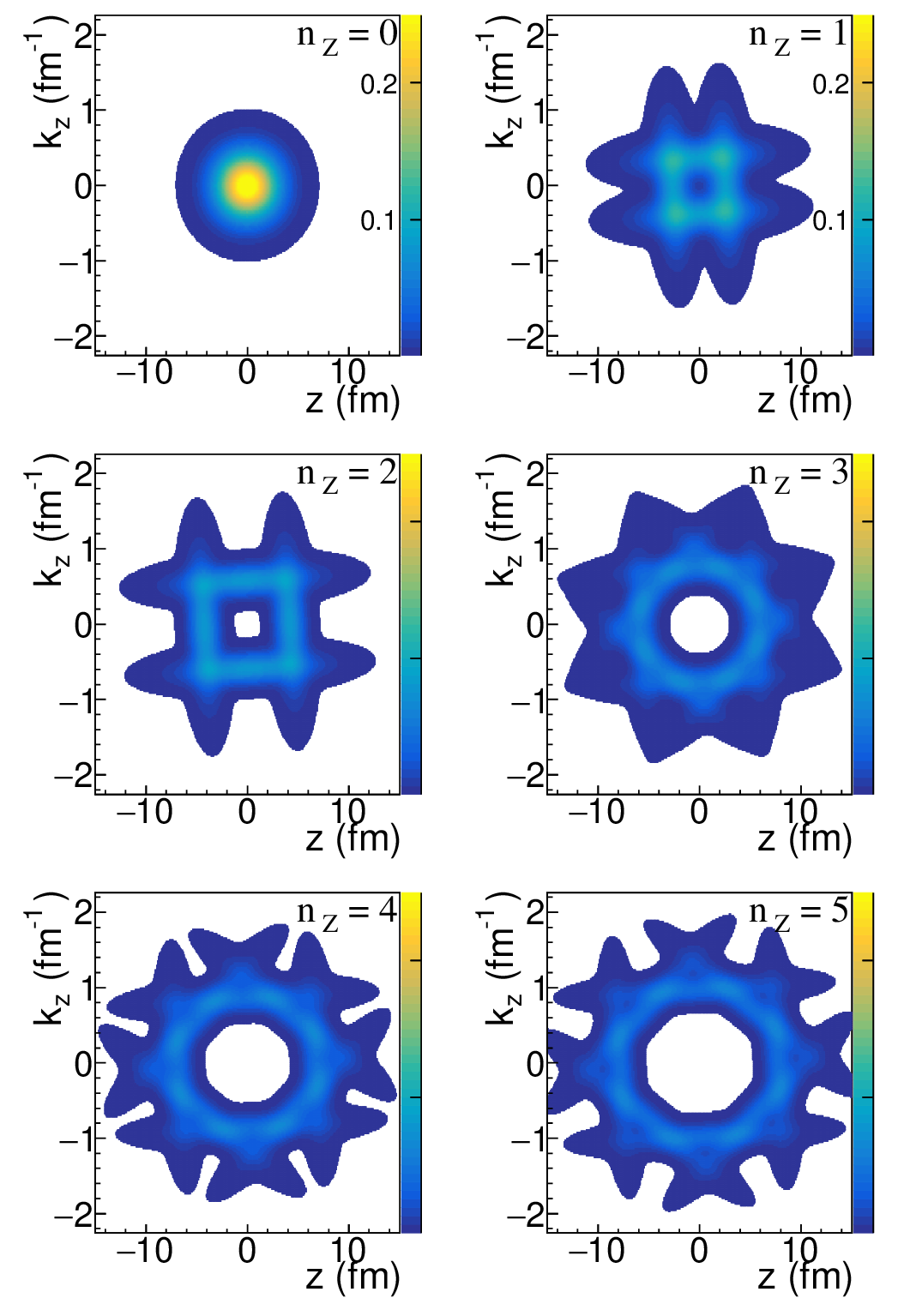}
\caption{(Color online.)
Each plot represents $P_{\lambda}(z,k_z)$, which is the contribution 
to the spatial density of a particular SP state $\lambda$ given by 
Eq.\ (\protect\ref{int}) corresponding to $n_z$ from 0 to 5 and 
$n_x=n_y=0$, projected onto the $(z,k_z)$ plane.
\label{splevel}}
\end{figure}

\subsection{\label{sec2a}The initial conditions}
The primal point in transport models is the initial condition: a trustworthy picture of the dynamics in heavy-ion collisions (HIC) must rely on correct static conditions of nuclear systems. 
In order to prepare nuclei in their ground state, the Hartree-Fock (HF) equation 
\begin{equation}\label{hf}
\ [ ~ \bf{h}, \rho \ ] = 0
\end{equation}
is solved through an iterative self-consistent procedure. 
Here, $\bf{h}$ is the one-body Hamiltonian ${\bf t}\,+\,{\bf U}(\rho)$, where ${\bf t}$ is the kinetic energy and ${\bf U}(\rho)$ the nuclear mean-field potential. 
In the present version of the model, we implemented an effective zero-range interaction of the Skyrme type, also accounting for asymmetry and surface terms:
\begin{eqnarray}
U_q & = & \frac{3}{4} t_0 \rho + \frac{(\sigma+2)}{16}t_3 \rho^{\sigma+1} \nonumber \\
 & - & \frac{\sigma t_3}{24}\left(x_3+\frac{1}{2}\right) \rho^{\sigma-1} \xi^2  \nonumber \\
   & -& \tau_{q}\frac{t_3}{12}\left(x_3+\frac{1}{2}\right) \rho^{\sigma} \xi  \nonumber \\
      & - & \tau_{q} \frac{1}{2} t_0 \left( x_0+\frac{1}{2}\right)\xi  \nonumber \\
         & - & \frac{1}{8} \left[ \frac{9}{4} t_1 - t_2 \left(x_2 + \frac{5}{4} \right) \right] \Delta\rho \nonumber \\
	  &+& \tau_{q} \frac{1}{16} \left[ 3 t_1 \left(x_1 + \frac{1}{2} \right)+ t_2 \left(x_2 + \frac{1}{2} \right) \right] \Delta\xi. \label{skyrme}
\end{eqnarray}
\noindent
%%% as described in Eq.\ (\ref{skyrme}). 
In this equation, $q$ represents the isospin, $\rho= \rho_n+\rho_p$, $\xi= \rho_n-\rho_p$, and $\tau_q=\pm 1$, the upper symbol holding for neutrons and the lower for protons. 
This implementation allows us to use any parametrization of this type. 
However, in this work for practical applications the parametrization of Ref.\ \cite{skt5}, hereafter Skt5, has been carried out.
Without entering into technical aspects we point out that the 
descriptions of initial states of nuclei are performed in three 
dimensions without any symmetry assumption. 
Nevertheless, an approximation is made at the level of the HF 
self-consistent procedure. 
In order to economize the numerical effort, the mean field is fitted 
with a one-dimensional harmonic well and extrapolated to the full 
three-dimensional space. 

The Wigner transform ($\mathcal{T}^W$) of dynamical densities is useful to obtain a classical-like representation of the dynamics in phase space. The corresponding contribution of a given CS $| \alpha \rangle$ to the one-body density matrix (\ref{ro1}) is given for simplicity in one dimension by the following expression:
\begin{eqnarray}
f_{\alpha}(x,k_x)=\mathcal{T}^W[ |\alpha\rangle \langle \alpha | ](x,k_x) =\frac{1}{2\pi\hbar\sqrt{\Delta}} \times \nonumber \\
 \exp \{  -\frac{1}{2 \Delta} \big[ \phi(x-\langle x \rangle)^2+\chi(k_x-\langle k_x \rangle)^2 \nonumber \\
  -2\sigma(x-\langle x \rangle)(k_x-\langle k_x \rangle) \big] \},
\end{eqnarray}
\noindent
with $ \Delta=\phi\chi-\sigma^2=\frac{1}{4}$ and $\phi=\langle(k_x-\langle k_x \rangle)^2 \rangle$. 
The Wigner transform of the one-body matrix density (\ref{ro1}) is then
%% \begin{equation}\label{int}
$$\rho(\vec r,\vec k)=\mathcal{T}^W[ \rho ](\vec r,\vec k)=\sum _{\lambda} \sum _i n_i^{\lambda}f_{\alpha_i^{\lambda}} (\vec r,\vec k) , $$
%% \end{equation}
where $f_{\alpha_i^{\lambda}}=f_{\alpha_{i_x}^{n_x}}(x,k_x)f_{\alpha_{i_y}^{n_y}}(y,k_y)f_{\alpha_{i_z}^{n_z}}(z,k_z)$.
Solving Eq.\ (\ref{hf}) with the above HF potential provides the 
complete set of SP wave functions describing the initial conditions. 
In Fig.\ \ref{splevel} we show the contributions of the SP states corresponding to $n_z$ from 0 to 5 to the projection onto the $(z,k_z)$ phase-space plane of $\rho(\vec r,\vec k)$, multiplied by a normalization factor
\begin{equation}\label{int}
P_{\lambda}(z,k_z)=\int\!\!\int \sum_i n_i^{\lambda}f_{\alpha_i^{\lambda}} 
(\vec r,\vec k) ~dx dk_x\,dy dk_y .
\end{equation}

\begin{figure}[tb]
\includegraphics[width=8.cm]{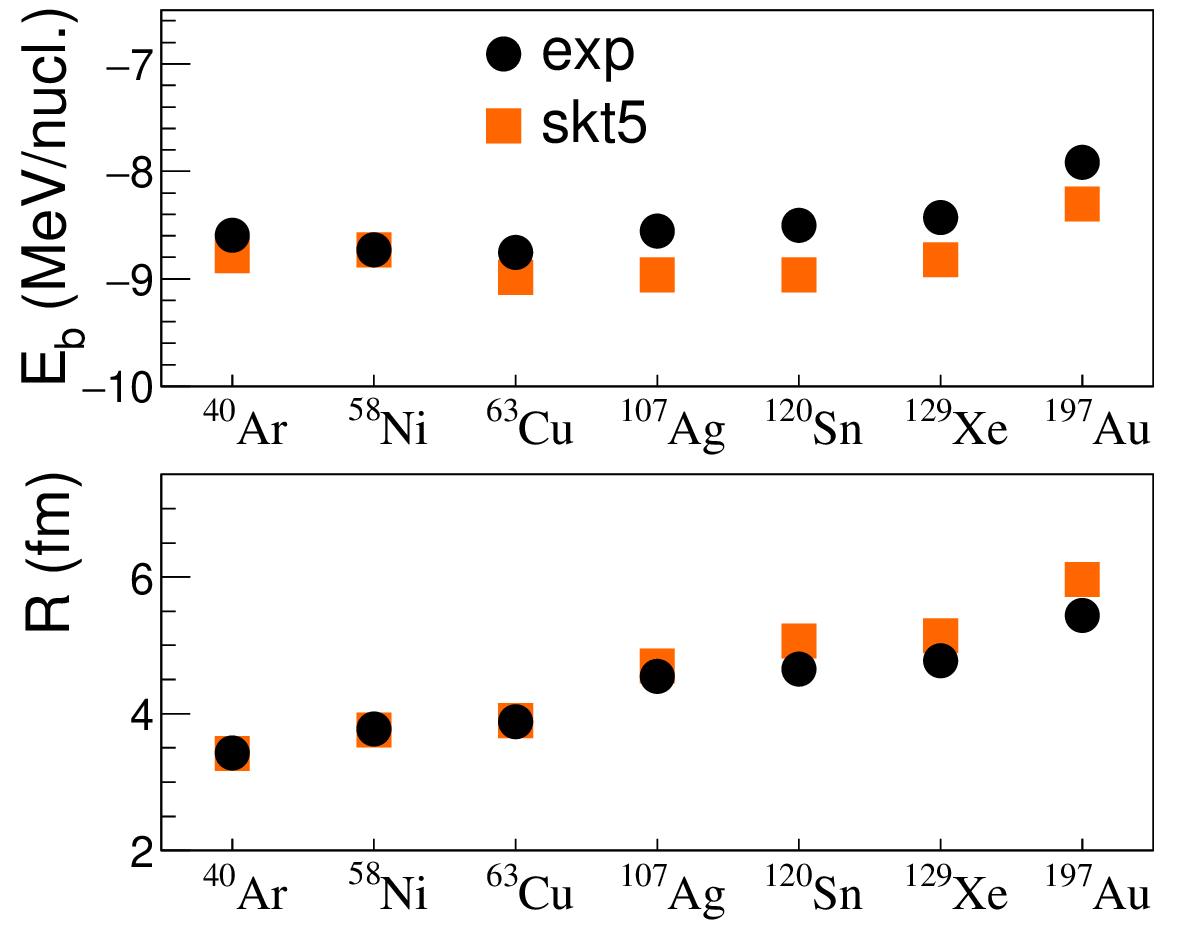}
\caption{(Color online.)
Ground state properties of nuclei addressed in this work (red 
squares) compared with experimental values (black dots): 
top panel, binding energy per nucleon, $E_{b}$; bottom panel, 
mean-square-radius $\langle R \rangle$.
\label{ebr}}
\end{figure}

Let us now present some ground-state properties of nuclei, typically studied in dynamical calculations, derived from the iterative resolution of the HF equation (\ref{hf}). 
In Fig.~\ref{ebr} the binding energy $E_b$ (top) and the mean square radius  $\langle R \rangle= \sqrt{\langle r^2 \rangle}$ (bottom) obtained with the mentioned Skt5 parametrization\ \cite{skt5} are represented and compared with the experimental values~\cite{expe}. We obtain an overall good agreement on a wide range of mass number $A$.
The relative errors are less than 5 and 10 percent, respectively, which is satisfactory in dynamical models for HIC at the concerned energies. These examples illustrate how the resulting static solutions constitute convenient initial conditions for a sizable spectrum of nuclei.

\begin{figure}[tb]
\includegraphics[width=7.cm]{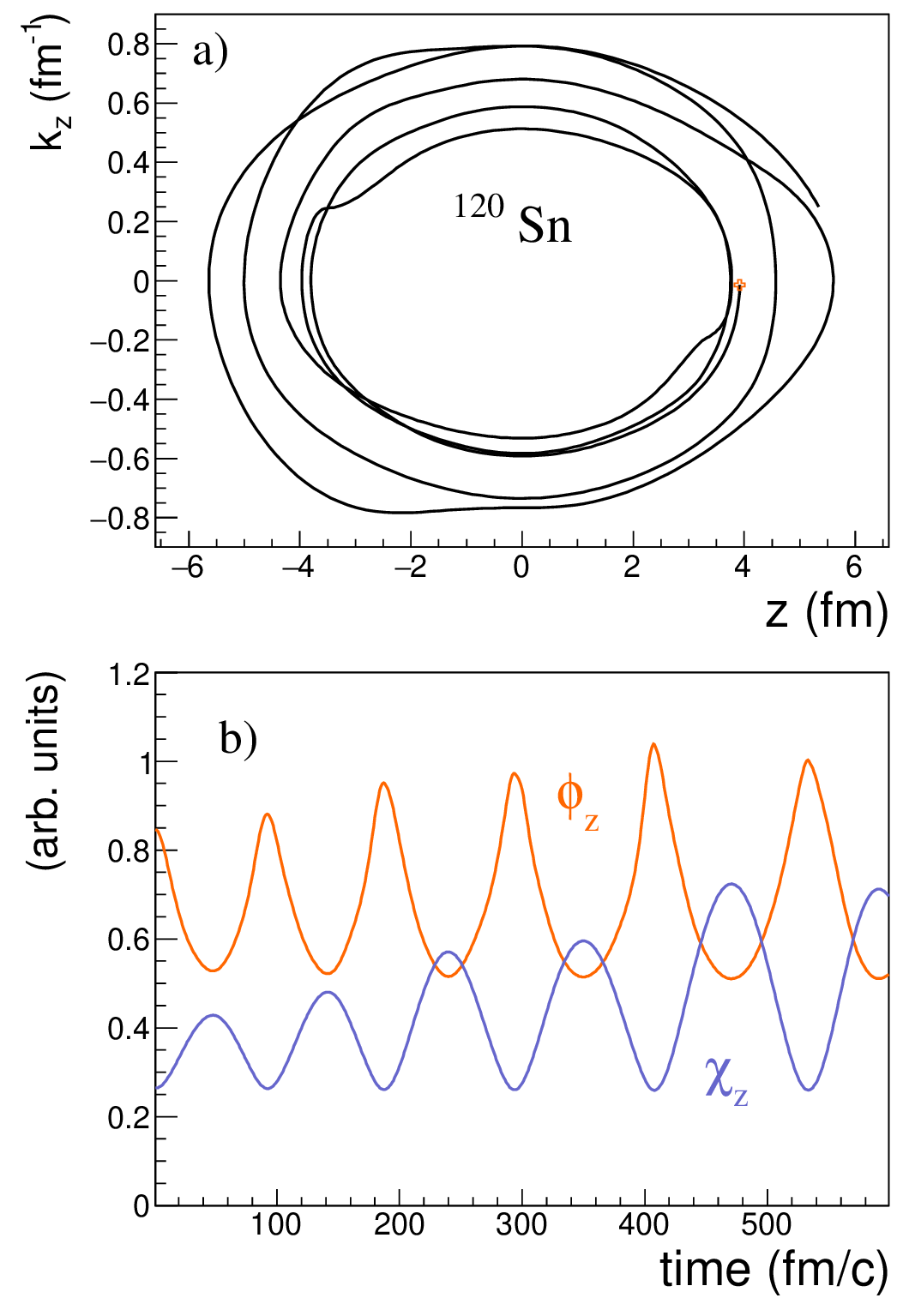}
\caption{(Color online.)
Time evolution of momentum (top) and spatial widths (bottom) 
of an isolated coherent state contributing to the $^{120}$Sn nucleus.
\label{trajecto}}
\end{figure}
\subsection{\label{sec2b}The mean-field evolution}
Let us now consider the equation of motion governing the slowly varying part of the
ETDHF equation (\ref{etdhf}),
\begin{equation}\label{tdhf}
i\hbar\dot{\rho}= \ [ ~ \bf{h}, \rho \ ]. 
\end{equation}
According to the adopted resolution scheme it is possible to derive from (\ref{tdhf}) a TDHF-like equation of motion for the $|\alpha_i^{\lambda}\rangle(t)$ states, of the form 
\begin{equation}
i\hbar | \dot{\alpha}_i^{\lambda}\rangle(t)=  {\bf h} |\alpha_i^{\lambda}\rangle(t) \label{alfadot}.
\end{equation}
\noindent
Equation (\ref{alfadot}) is solved by a variational principle giving the following equations of motion for first and second moments in phase space:
\begin{eqnarray}\label{dotx}
\dot{\langle x \rangle}&=&\frac{\hbar}{m}\langle k_x \rangle , \\
\label{dotxk}
\dot{\langle k_x \rangle}&=&-\frac{1}{\hbar}\frac{\partial}{\partial \langle x \rangle} {\cal{U}} , \\
\label{dotki}
\dot{\chi}&=&\frac{4 \hbar}{m}\gamma \chi , \\
\label{dotga}
\dot{\gamma}&=&\frac{\hbar} {8 m}\frac{1}{ \chi^2} - \frac{2 \hbar} {m}\gamma^2-\frac{1}{\hbar}\frac{\partial}{\partial \chi} {\cal{U}} , 
\end{eqnarray}  
\noindent
where $\gamma=\frac{\sigma}{2\chi}$ and ${\cal{U}}=\langle \alpha|U|\alpha \rangle$.
The time integration of the system of equations of motion of CSs is
carried out through the predictor-corrector method of second order, a
procedure which is appreciably faster than, e.g., a Runge-Kutta method
of the same level of numerical accuracy.
The typical time step in the dynamical calculation is 1 fm/$c$.

Figure \ref{trajecto} shows the evolution of first and second moments of an isolated CS contributing to the description of the $^{120}$Sn nucleus, followed up to several hundreds of fm/$c$. On top of this figure is the trajectory of the centroid in the $(x,k_x)$ plane, the initial time being tagged by a small open circle. On the bottom are the widths $\chi$ and $\phi$ corresponding, respectively, to $x$ and $k_x$ coordinates as a function of time. Since this particular CS remains bound during its evolution the widths oscillate in a nearly harmonic-modulated way. This behavior is due to the variation of the CS orientation all along the trajectory and to the conservation of the correlation relation $\Delta=1/4$. 

Let us concentrate now on the dynamical evolution of a nuclear system in the limit of vanishing binary collisions. This would be the case in HIC at low incident energies, where two-body collisions are suppressed as a consequence of the Pauli blocking. The present approach has not been designed to work at extremely low energies; nevertheless, it is convenient to control the performance of the model in a pure mean-field evolution in order to ensure the correct transition from the collisionless to the collision regime.

\begin{figure}[tb]
\includegraphics[width=7.cm]{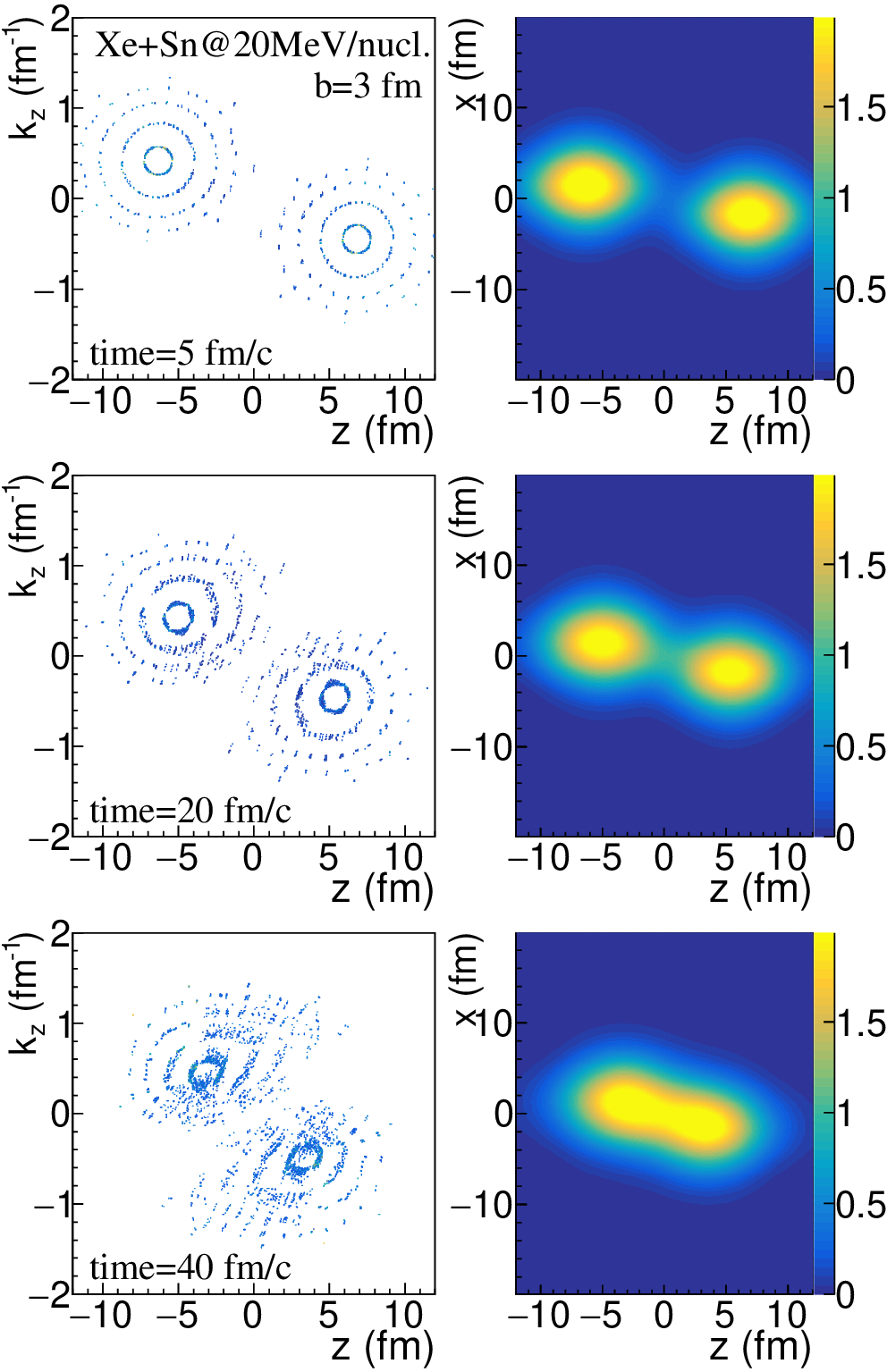}
\caption{(Color online.)
The complete set of coherent-state centroids of the 
$^{129}$Xe+$^{120}$Sn collision at 20 MeV/nucleon and $b\!=\!3$ fm 
projected at 
selected times onto the $(z,k_z)$ phase-space plane (left column) 
and the density-profile contours projected onto the reaction plane 
$(z,x)$ in the configuration space (right column).
The $z$ axis is along the incident beam.
The color palette is normalized to correspond to the number of 
system nucleons when integrated over the $(z,k_z)$ plane.
\label{centrost}}
\end{figure}

In order to illustrate the above delineated mean-field description,
Eqs.\ (\ref{dotx}) to (\ref{dotga}) are solved for some typical 
nuclear systems in the energy range we are concerned with.
The time evolution of the complete set of CS centroids projected on the 
$(z,k_z)$ phase-space plane (left column) and the configuration space 
density projected onto the reaction plane $(z,x)$ (right column)
is represented in Fig.\ \ref{centrost} for the $^{129}$Xe+$^{120}$Sn collision at 20 MeV/nucleon and impact parameter $b\!=\!3$ fm. 
From the left column one infers that the 
centroids are initially located in shells corresponding to different single-particle energy levels and, as time proceeds, they move in almost closed orbits which are progressively deformed by the dynamics. These phase-space trajectories are the projection on one dimension of six-dimension orbits, the  evolution of which generates shape deformations and orbit crossings.

\subsection{\label{sec2c}The dissipative behavior}
In the ETDHF framework the mean-field flow provided by Eqs.\ (\ref{dotx})--(\ref{dotga}) is coupled to a master equation for CS occupation numbers describing residual correlations. Indeed, starting from the CS expansion (\ref{ro1}) the collision term ${\bf I}(\rho)$ ruling the evolution of SP occupation numbers $\eta_{\lambda}$ can be written in the form of a gain-minus-loss equation in terms of CS occupation numbers $n_m^{\mu}$  
\begin{eqnarray}
\dot{\eta}_{\beta} & = & \sum_{\gamma \lambda \mu}  \mathcal{W}_{\beta \gamma \lambda \mu} \sum_{ijlm} \nonumber \\
 & & \times \big[(1-\Theta_i)n_i^{\beta}(1-\Theta_j)n_j^{\gamma} \Theta_l n_l^{\lambda}  \Theta_m n_m^{\mu}  \nonumber \\
 & & - \Theta_i n_i^{\beta}\Theta_j n_j^{\gamma}  (1-\Theta_l) n_l^{\lambda}  (1-\Theta_m) n_m^{\mu} \big] \label{nmun} , 
\end{eqnarray}
\noindent
where $\Theta_m\!=\!1$ if $\alpha_m^{\mu}$ takes part in the SP state $\varphi_{\mu}$ expansion or 0 otherwise, so that
$$\eta_{\mu}=\sum_{m=1}^{m_\mu} n_m^{\mu} \Theta_m.$$
The $\mathcal{W}_{\beta \gamma \lambda \mu}$ are the transition probabilities obtained in the Born and in the Markovian approximations and in the weak interaction limit. In transport models they are usually formulated in terms of the nucleon-nucleon cross-section
\begin{eqnarray}\label{kernel}
\mathcal{W}_{\beta\gamma\lambda\mu} & = & w \frac{d\sigma_{NN}}{d\Omega}(\vec k_{\beta},\vec k_{\gamma},\vec k_{\lambda},\vec k_{\mu})
\delta(\vec k_{\beta}+\vec k_{\gamma}-\vec k_{\lambda}-\vec k_{\mu}) \nonumber \\
     & \times & \delta(\varepsilon_{\beta}+\varepsilon_{\gamma}-\varepsilon_{\lambda}-\varepsilon_{\mu}),
\end{eqnarray}
\noindent
where $\vec k_{\alpha}$ and $\varepsilon_{\alpha}$ are SP momenta and
energies and $w$ is a constant factor.

Taking advantage of CS dynamical properties, the resolution scheme adopted for the collision term is based on the philosophy of test-particles methods \cite{testp,lv,dywan}. In CS diffusions a normalized nucleon-nucleon cross section is implemented: $\frac{2\sigma}{(|c_i|^2+|c_j|^2)}$, $|c_i|^2$ being the weights of the colliding CS in Eq.\ (\ref{varfi}). The Pauli principle is ensured by suppressing all diffusions inside the same phase-space elementary volume of size $\hbar^3$. This requirement can be simply related to a minimum overlap condition of the scattered CS with all the others. 
In Ref.\ \cite {dywan} a preliminary resolution scheme of the ETDHF equation was proposed and different aspects of the description were analyzed. Applications to HIC at intermediate energies \cite{dywan2} and to nucleon induced collisions \cite{dywan22} have been performed.  An extension of the model  adapted to the description of the outer layers of neutron star crusts has also been developed \cite{dywan3}. Since then, the modelization techniques have evolved improving the quality of the corresponding phase-space representation while optimizing the numerical framework\ \cite{these}. In this sense the choice  of the representation basis is determinant because it guarantees the Pauli exclusion principle at all times. The role of wavelets in this description is restricted to the multiresolution analysis of SP wave functions in the initial conditions in terms of a set of CSs, the latter being the essential ingredient in this approach. 

The expression of transition probabilities (\ref{kernel}) reflects 
energy and momentum conservation of individual collisions. 
At the macroscopic level there is no explicit energy conservation 
constraint. 
As a consequence, the dynamical evolution of a nuclear system is 
subjected to fluctuations of the overall mean energy over the course 
of time, which is system dependent. 
A systematic study of the total energy of the studied systems allows 
us to assert that the corresponding mean uncertainty is of the order 
of a few percent for several hundred fm/$c$.

\begin{figure}[tb]
\includegraphics[width=7.cm]{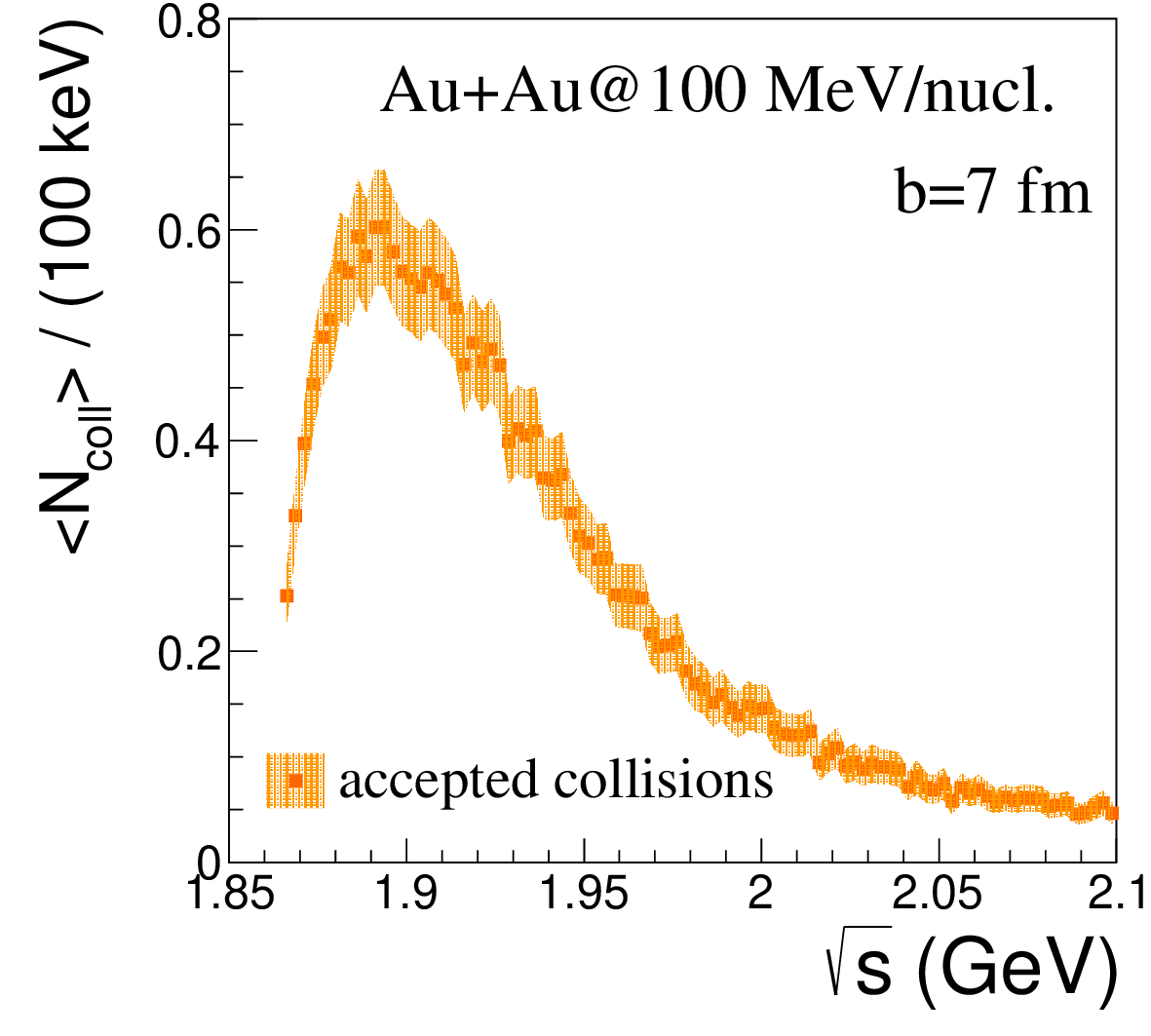}
\caption{(Color online.)
The collision rate integrated over 140 fm/$c$ for the 
$^{197}$Au+$^{197}$Au reaction at 100 MeV/nucleon and $b\!=\!7$ fm 
as a function of c.m.\ energy $\sqrt{s}$.
The errors are statistical, i.e., they are the square root of the 
number of accepted collisions.
\label{coll} }
\end{figure}

As an usual test of the actual numerical implementation, versus other 
existing approaches compared among themselves in Ref.\ \cite{xu} (cf.\ 
Figs.\ 7 and 8 in\ \cite{xu}), in Fig.\ \ref{coll} we show
the number of accepted (``successful'') collisions per 100-keV 
bin and per nucleon as a function of the center-of-mass (c.m.) energy 
$\sqrt{s}$ for the case of the Au\,+\,Au reaction at 100 MeV/nucleon 
and $b\,=\,7$ fm.
The observed behavior is entirely compatible with the other theoretical 
approaches to HIC compared in\ \cite{xu}.
Let us finally mention that, besides the CS basis, the implementation 
of other kinds of expansion functions in order to improve the numerical 
treatment is an open issue.     

\subsection{\label{sec2d}Fluctuations in dissipative processes}
In order to estimate the dispersions of measured observables around their mean values a procedure for untangling many-body information from the ETDHF approach is mandatory. To this end let us consider the contingency of a model to provide a complete many-body description of the system through the $N$-body density matrix.
Even if the system could ideally be prepared in a non-correlated initial state and, accordingly, be described by a unique Slater determinant (SD) of individual nucleon states, the dynamical correlations would generate transitions between different states in a broad many-body space. Since knowledge of the evolution of the relative phase shifts is unattainable, the most general $N$-body density matrix we could construct is of the form
\begin{eqnarray}\label{roN}
\rho^{(N)}=\sum_k^K p_k |\Psi_{k} \rangle \langle \Psi_{k}| , 
\end{eqnarray}
where $\lbrace \Psi_{k} \rbrace$ is a  set of $K$ nucleonic SDs (NSDs) contributing to an incoherent admixture of states with probabilities $p_k$. According to the CS expansion of single particle states $\lbrace \varphi_{\lambda}\rbrace$, Eq.\ (\ref{varfi}), the many-body density (\ref{roN}) is shown to be 
\begin{eqnarray}\label{roNk}
\rho^N=\sum_{k=1}^K p_k \sum_{q=1}^{\mathcal{N}_q^{(k)}}  |c^{(k)}|^2 | \Pi_q^{(k)}\rangle \langle \Pi_q^{(k)}| , 
\end{eqnarray}
where $k$ corresponds to a set of nucleon quantum numbers $k=\lbrace \lambda_1, ... , \lambda_N \rbrace$, $q=\lbrace i_1,i_2,...,i_{n}, ... ,i_N\rbrace$ is a collection of CS labels participating in the decomposition of nucleonic states in a given $\Psi_k$ and, for $\mathcal{A}$ being the antisymmetrization operator, $| \Pi_q^{(k)}\rangle=\mathcal{A}| \alpha_{i_1}^{\lambda_1} \alpha_{i_2}^{\lambda_2} ... \alpha_{i_N}^{\lambda_N}\rangle$ is a SD of coherent states (CSSD). In  Eq.\ (\ref{roNk})  $K$ is the total number of NSDs and, since $m_{\lambda}$ is the number of CSs contributing to a nucleon state $| \varphi_{\lambda}\rangle$, the product $\mathcal{N}_q^{(k)}=m_{\lambda_1} \times ... \times m_{\lambda_N}$ represents the total number of different CSSDs for a given $k$ value. Since CS coefficients are identical for a given SP state $\lambda$, then in Eq.\ (\ref{roNk}) coefficients 
$c^{(k)}=c^{\lambda_1}\times c^{\lambda_2}\times . . . \times c^{\lambda_N}$ only depend on label $k$.

Assembling all CSSDs generated by the complete set of NSDs contributing to $\rho^N$, one can simply write
\begin{equation}\label{roNi}
\rho^N=\sum_I^{\mathcal{N}} \omega_I | \Pi_I\rangle \langle \Pi_I| ,
\end{equation}
where $\mathcal{N}=\sum_k \mathcal{N}_q^{(k)}$ is the dimension of the complete set of CSSDs and the definition of the weights $\omega_I$ follows immediately from Eqs.\ (\ref{roNk}) and (\ref{roNi}). Accordingly, the one-body density matrix is 
\begin{eqnarray}\label{ro2}
\rho=NTr_{2 ... N}  \lbrace \rho^N \rbrace =\sum_I \omega_I \rho_I ,
\end{eqnarray}
where the symbol $Tr_{2 ... N}$ represents the trace over the degrees of freedom of 2, 3, ..., $N$ particles. Here 
$$\rho_I=N Tr_{2...3}\lbrace | \Pi_I\rangle \langle \Pi_I | \rbrace$$ 
is the one-body density matrix associated with the CSSD $| \Pi_I\rangle$.  

The average value of any one-body observable ${\bf B}=\sum_{n=1}^N {\bf b}(n)$ can be calculated as
\begin{equation}\label{ave}
\langle {\bf B} \rangle=\sum_{I=1}^{\mathcal{N}} \omega_I Tr \lbrace \rho_I {\bf b} \rbrace=\sum_{I,i} \omega_I \langle \alpha_i^I | {\bf b} |\alpha_i^I\rangle ,
\end{equation}
where $i$ spans all CS labels in the corresponding CSSD $|\Pi_I\rangle$.
For a finite NSD set, the probabilities of which are unknown, the sums in Eq.\ (\ref{ave}) represent a huge amount of contributions. In order to compute this equation, a simple sampling of CSSD with uniform density random variable is then performed. 
In this way Eq.\ (\ref{ave}) results in
\begin{eqnarray}
\langle {\bf B} \rangle &\simeq& \frac{1}{N_{sd}}\sum_{l=1}^{N_{sd}} b_l(N) \label{avesd}, \\
b_l(N)&=&\langle \alpha_{i_1}^l |{\bf b} |\alpha_{i_1}^l \rangle +...+\langle \alpha_{i_N}^l |{\bf b} |\alpha_{i_N}^l \rangle \label{avesd1},
\end{eqnarray}
with $N_{sd}$ the number of SD samples. The quantities $\langle \alpha_{i}^l |{\bf b} |\alpha_{i}^l \rangle$ are the mean values of the ${\bf b}$ observable for individual coherent states $|\alpha_{i}^l \rangle$. On the other side $b_l(N)$ represent the corresponding SD average values spreading out from $\langle {\bf B} \rangle$ with a given width. Although we will be only concerned here with one-body type observables, a similar treatment can be shaped to the estimation of many-body observables involving reduced density matrices.
 
The above considerations endorse the fact that the one-body density matrix can be viewed as a microcanonical ensemble average of microscopic $N$-body configurations constituted by individual CSSD buildup from ETDHF solutions. These configurations, which may be called ``events'', by construction conserve the overall mean values of observables, respect the Pauli principle, and, otherwise, they correspond to the least biased $N$-body description compatible with the one- and two-body information contained in the ETDHF approach. 
Let us finally mention the fact that for the simulation measure having the sense of a genuine one-body quantity, nucleon aggregates which should not be experimentally detected must not be computed in (\ref{avesd}). The selection of relevant events is performed through a cluster-recognition algorithm, which in this work has been adapted from Ref.\ \cite{cluster}.

To illustrate the dispersion of $N$-body configurations around the average values, we considered
the time evolution of the energy isotropy ratio $E_{\rm iso}$ defined 
as the quotient between the mean transverse kinetic energy 
$E_{\perp}$ and the mean longitudinal kinetic energy $E_{\parallel}$
\begin{equation}
E_{\rm iso} = \frac{\langle E_{\perp}\rangle}{2\,\langle E_{\parallel} \rangle}.
\label{r_e1}
\end{equation}
\begin{figure}[tb]
\includegraphics[width=7.cm]{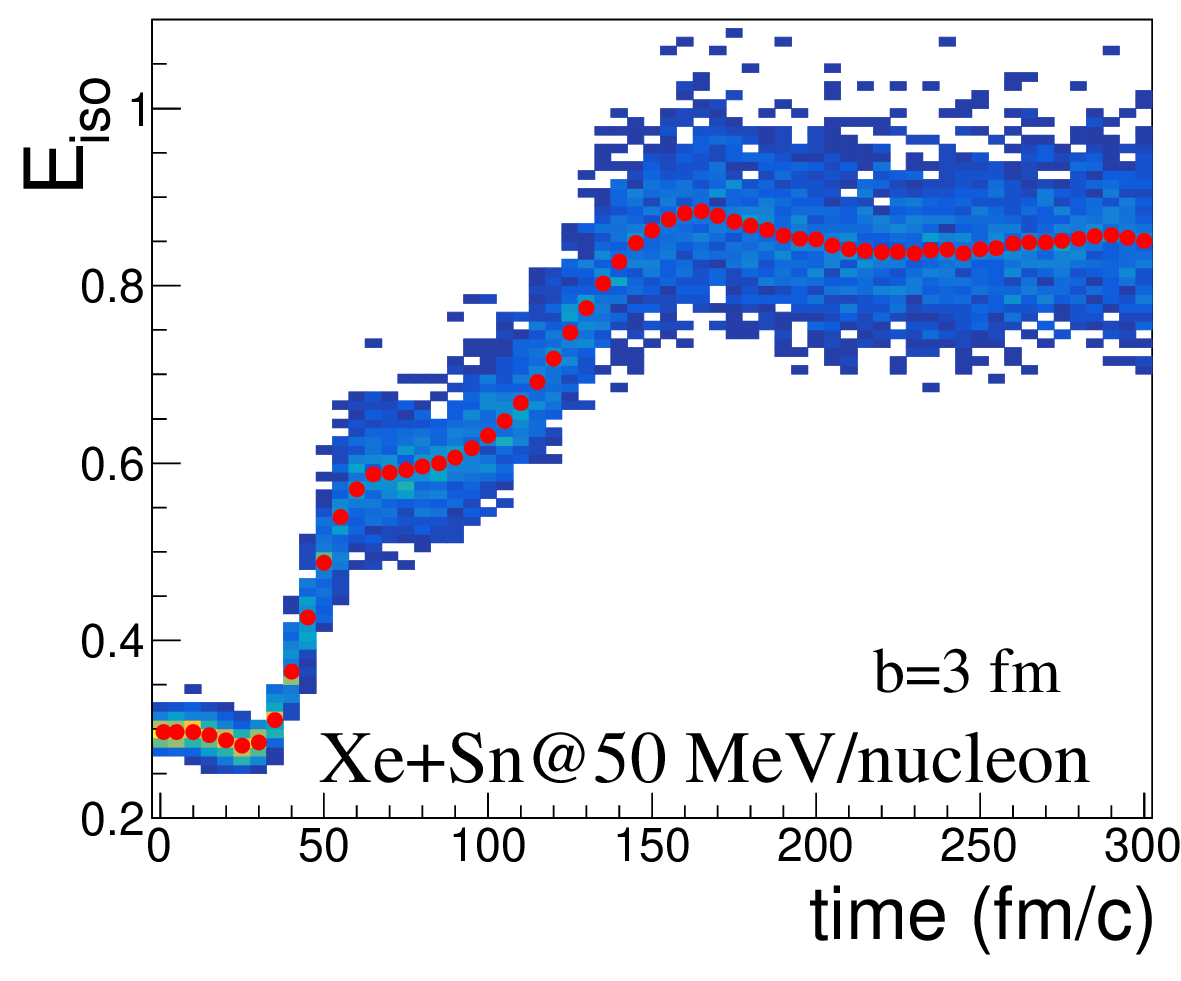}
\caption{(Color online.)
The isotropy ratio $E_{\rm iso}$ as given by Eq.\ 
(\protect\ref{r_e1}) for the $^{129}$Xe\,+\,$^{120}$Sn reaction 
at 50 MeV/nucleon and $b$\,=\,3 fm as a function of time.
The red dots show the results of the dissipative evolution 
(Eq.\ (\protect\ref{etdhf})).
The bluish background is due to the dispersion over 1\,000 CSSDs 
according to Eq.\ (\protect\ref{avesd1}).
\label{slat}}
\end{figure}
\noindent
Thus defined, this observable is of a one-body type.
It has been introduced experimentally\ \cite{lehaut,lopez} 
and widely studied theoretically\ \cite{iqmd1,su13,iqmd2,zhang,zhao,bonnet,li18,bas16} 
and will be discussed in more detail in Sec.\ \ref{sec3}. 
In the example shown in Fig.\ \ref{slat} the full ETDHF 
calculation has been performed for the $^{129}$Xe on $^{120}$Sn collision at 
50 MeV incident energy per nucleon $E_{\rm inc}$ for an impact 
parameter $b$ of 3 fm 
and the free-space nucleon-nucleon ($NN$) cross section. 
The average $E_{\rm iso}$ values (red dots) firstly strongly
increase with time and quickly stagnate tending towards the value
observed experimentally.
The corresponding dispersion resulting from a sample of 1\,000 CSSDS 
is shown in blue, where the hue from lighter to darker blue denotes 
decreasing statistics of CSSDs. 

After presenting the main ingredients and features of the model in 
the next section, we 
will focus on dynamical aspects of HIC through the description of 
one-body observables that strongly depend on the two-body dissipation.

\section{\label{sec3}Stopping observables}
Several experimental observations are sensitive to the 
nuclear stopping power\ \cite{lehaut,lopez,fopi,colin}.
These observables have recently been abundantly investigated 
by a number of model-studies of HIC (IQMD\ 
\cite{iqmd1,su13,iqmd2,zhang}, AMD\ \cite{zhao}, 
SMF\ \cite{bonnet}, UrQMD\ \cite{li18}, LV\ \cite{bas16}, and BUU\ 
\cite{bar19}) in order to constrain the elastic $NN$ 
cross section inherent to these models.
In Fig.\ \ref{slat} we show the power of the present 
approach in extending the model via a procedure which traces back 
statistical fluctuations in a mesoscopic quantal system.
However, at this stage of the model, as already mentioned, 
the present paper is intended to illustrate the immanent features 
of the model and not to carry out a quantitative comparison with 
experimental results.

To this end we chose to confront qualitatively our model with 
two stopping-sensitive experimental observables: the isotropy 
ratio applied to hydrogenlike reaction products \ \cite{lopez}
and the linear momentum transfer (LMT)\ \cite{colin}.
Our choice is additionally motivated by the fact that, from the 
experimental point of view, these two observables are very different. 
LMT is simple to extract for a given class of experimental events 
whereas $R_E$ is not only highly sensitive to the details of event 
selection but also appreciably depends on the choice of the 
representative particles to be studied within selected events.

\begin{figure}[tb]
\includegraphics[width=\columnwidth]{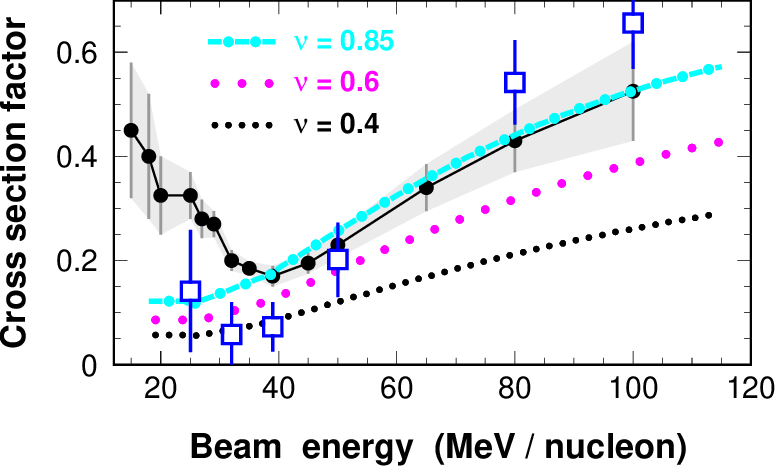}
\caption{(Color online.)
The global cross-section reduction factor ${\cal F}$ of Eq.\ 
(\protect\ref{fct_f}) for $\nu\!=\!0.85$, 0.6, and 0.4, 
respectively as a function of $E_{\rm inc}$.
The heavy dots and the background gray zone are ${\cal F}$ values 
and attributed uncertainties extracted from experiment\ 
\protect\cite{lopez}.
The open squares stand for the ${\cal F}$ values deduced in 
the LV semiclassical analysis of the observable $R_E$\ 
\protect\cite{bas16}, which used the Zamick parametrization\ 
of the nuclear equation of state\ \protect\cite{zamick}.
\label{sig_f}}
\end{figure}

It has been demonstrated in all previous investigations\
\cite{iqmd1,su13,iqmd2,zhang,zhao,bonnet,li18,bas16,bar19}
that, to agree with experimental results on stopping, the cross section
responsible for the two-body dissipation in Eq.\ (\ref{kernel}) 
should be strongly reduced relative to the free-space 
$NN$ cross section $\sigma^{\rm free}_{\rm NN}$ in order to account 
for its in-medium modification. 
To that aim we use the prescription of Refs.\ \cite{dan02} 
which, besides reducing $\sigma^{\rm free}_{\rm NN}$, strongly dumps 
its dependence on energy and introduces the dependence on density 
$\rho$:
\begin{eqnarray}
\nonumber
\sigma^{\rm med}_{\rm NN} & = & \sigma_0 ~ \tanh (\sigma^{\rm free} / \sigma_0), \\
\sigma_0 & = & \nu / \rho^{2/3} ,
\label{pnu}
\end{eqnarray}
\noindent
where the factor $\sigma_0$ is motivated by the assumption that the geometric 
cross-section radius should not exceed the interparticle distance 
\cite{dan02}.
The recommended value for the parameter $\nu$ has been 0.85\ 
\cite{dan02}.
For $E_{\rm inc}$ below 100 MeV/nucleon in the recent publication 
it has been suggested that it is more appropriate to use 
$\nu\!=\!0.4-0.6$\ \cite{bar19}.
We adopt both of the above recommended parameter values and add to our 
simulations those with the unmodified free $NN$ cross section 
$\sigma^{\rm free}_{\rm NN}$.
For $\sigma^{\rm free}_{\rm NN}$ we take the phenomenological 
parameterization by Chen et al.\ \cite{chen68}, which is based on the 
empirical isospin and energy dependence of the free $NN$ scattering.
To enable the comparison with the global in-medium modification factor 
${\cal F}$ extracted in\ \cite{lopez}, 
\begin{equation} 
{\cal F} = \frac{\sigma^{\rm med}_{\rm NN}}{\sigma^{\rm free}_{\rm NN}} 
\label{fct_f}
\end{equation}
\noindent
we adopt a constant value for the cross section of Eq.\ (\ref{pnu}) 
which, as mentioned above, washes out the dependence on $E_{\rm inc}$.
For $\nu\!=\!0.85$, $\sigma^{\rm med}_{\rm NN}$ varies from 27.7 
to 25.6 mb at energies between $E_{\rm inc}\!=\!17$ and 115 MeV/nucleon, 
while for $\nu\!=\!0.6$ it varies from 19.55 to 19.15 mb, and for 
$\nu\!=\!0.4$ it goes from 13.03 to 13.01 mb.
As in\ \cite{lopez} we take $\rho\!=\!0.17$ fm$^{-3}$.
The corresponding ${\cal F}$ values are shown in Fig.\ \ref{sig_f}.
One plausibly infers from this figure that $\nu\!=\!0.6$ and especially 
$\nu\!=\!0.4$ should be excluded according to the analyses of Ref.\ \cite{lopez}.
Each simulation result and the associated uncertainty shown in Fig.\ 
\ref{iso} (Figs.\ \ref{lmt} and \ref{lmt_1}) are due to 1\,000 (100) 
events generated through the mentioned prescription in Sec.\ \ref{sec2d}
with the cluster-recognition algorithm\ \cite{cluster}.
An analysis of this kind has been carried out previously in the 
framework of the semiclassical Landau-Vlasov (LV) approach\ 
\cite{bas16}, which has been shown to provide a good description 
of the dynamics of HIC at intermediate energies\ \cite{lv_g}.
For the sake of comparison with semiclassical approaches, in Fig.\ 
\ref{sig_f} (open squares) are the ${\cal F}$ values obtained 
in the mentioned analysis of the observable $R_E$ with the LV model\ 
\cite{bas16} and a local nuclear potential in the Zamick 
parametrization\ \cite{zamick}.
Taking into account the uncertainties on the deduced ${\cal F}$ values 
in both simulation and data, one may argue that a satisfactory agreement 
prevails mainly at the highest $E_{\rm inc}$.
Still, the general steepness of the $E_{\rm inc}$ dependence of 
${\cal F}$ is not well reproduced.

\subsection{\label{sec3a}Energy isotropy ratio}
\begin{figure}[tb]
\includegraphics[width=8.cm]{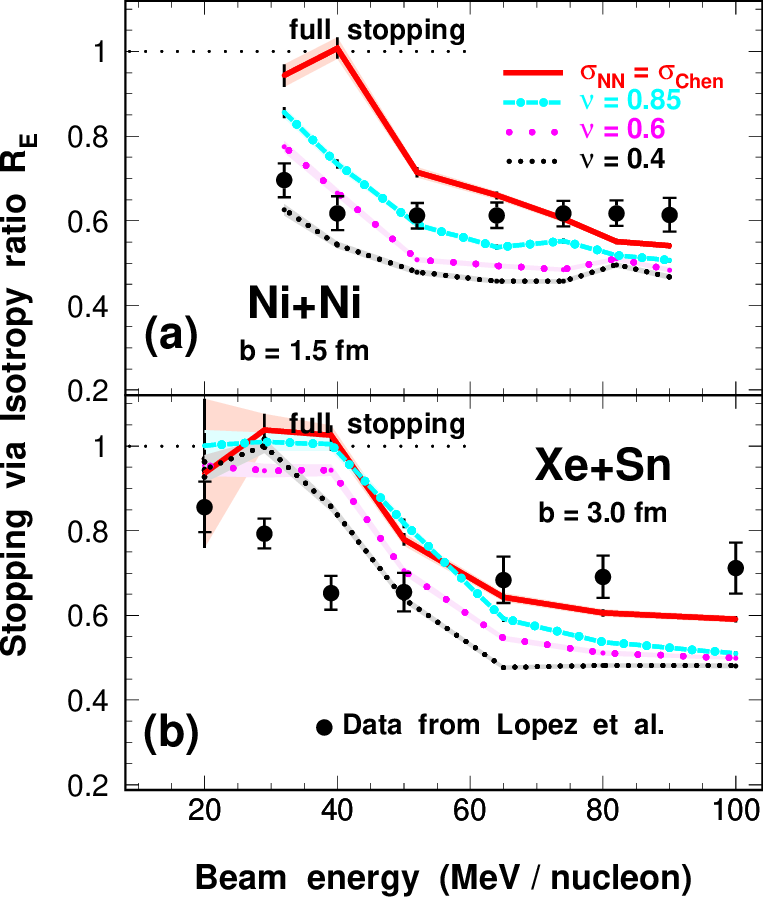}
\caption{(Color online.)
Isotropy ratio $R_E$ of Eq.\ (\protect\ref{r_e}) for (a) 
$^{58}$Ni\,+\,$^{58}$Ni and (b) $^{129}$Xe\,+\,$^{120}$Sn 
reactions as a function of $E_{\rm inc}$.
Lines show simulation results with varied parameter $\nu$
of Eq.\ (\protect\ref{pnu}).
The value of impact parameter $b$ is chosen according to\ 
\protect\cite{bas16}.
The dimmed background stands for the estimated errors.
Symbols represent experimental data\ \protect\cite{lopez}.
\label{iso}}
\end{figure}

The isotropy ratio $R_E$ measures the magnitude of energy transfers 
between longitudinal and transverse directions in a reaction.
This observable, defined in Eq.\ (\ref{r_e1}), can be computed in a 
simulation on the same footing as in an experiment:
\begin{equation}
R_E = \frac{\sum_i E_{\perp}^{i}}{2\,\sum_i E_{\parallel}^{i}} .
\label{r_e}
\end{equation}
\noindent
where $E_{\parallel}^i$ and $E_{\perp}^i$ are the longitudinal and 
transverse energy components of the $i^{\rm th}$ reaction ejectile. 
In the study of several mass-symmetric reaction systems\ \cite{lopez} 
the summation in Eq.\ (\ref{r_e}) runs over hydrogenlike products.
In Fig.\ \ref{iso} the corresponding $R_E$ values for the reactions
$^{58}$Ni+$^{58}$Ni and $^{129}$Xe+$^{120}$Sn are shown in the 
upper and lower panels, respectively.
The heavy dots are from the experiment\ \cite{lopez} and the various line
types are due to the simulation.
According to the results on the $b$ distribution of the 
$^{129}$Xe+$^{120}$Sn collisions corresponding to the most
violent subset of 0.02\,$\sigma_R$,\footnote{%
  2.0\,\% of total reaction cross section $\sigma_R$ is the
  geometrical cross-section equivalence for the fraction of the
  most central events selected in the experimental study of
  isotropy ratio in Ref.\ \protect\cite{lopez}.}
shown in Fig.\ 4 of Ref.\ \cite{bas16} the most contributing 
collisions are those with $b\!=\!3$ fm.
This is why in the present qualitative study for the 
$^{129}$Xe+$^{120}$Sn reaction we chose $b\!=\!3$ fm and for the 
$^{58}$Ni\,+\,$^{58}$Ni one $b\!=\!1.5$ fm.

To strictly follow the experimental procedure\ \cite{lopez}, the 
reported $R_E^{\rm theo}$ values are obtained by summing-up 
the contribution of the energy components of $Z\!=\!1$ particles 
from all the 1\,000 generated events.
However, to extract the uncertainty over the so-obtained 
$R_E^{\rm theo}$ value (see dimmed background in Fig.\ \ref{iso}) 
a parallel analysis based on event-by-event application of Eq.\ 
(\ref{r_e}) was performed too.

The experimentally deduced values $R_E^{\rm exp}$ firstly diminish 
with increasing $E_{\rm inc}$ up to about the Fermi energy $E_F$. 
Above this energy $R_E^{\rm exp}$ is nearly constant or very weakly 
rising with $E_{\rm inc}$.
In the simulation the overall behavior of $R_E^{\rm theo}$ is 
similar although the break in the $R_E^{\rm theo}$ slope occurs at 
10 to 20 MeV/nucleon higher $E_{\rm inc}$.
Nevertheless, the dependence on the parameter $\nu$ is very 
ambiguous to the point that at the highest $E_{\rm inc}$ the 
claim of the need, for the in-medium reduction of $\sigma_{\rm NN}$ 
is uncertain.
However, for the lighter $^{58}$Ni+$^{58}$Ni reaction [see panel (a) in Fig.\ 
\ref{iso}] and by considering the full range of $E_{\rm inc}$ one 
may conjecture that the $\nu\!=\!0.85$ curve is the closest to the 
ensemble of experimental $R_E^{\rm exp}$ values.
In contrast, for the heavier Xe+Sn reaction [see Fig.\ 
\ref{iso}(b)] the discrepancy between $R_E^{\rm theo}$ and 
$R_E^{\rm exp}$ is present at all but the lowest and intermediate 
studied $E_{\rm inc}$.
For $E_{\rm inc}\!\leq\!E_F$ whatever is the value of $\nu$ the 
simulation indicates the fusionlike behavior characterized by 
$R_E\!\gtrsim\!0.9$.
It is worth noting that for the Xe\,+\,Sn reaction incomplete 
fusion is present for $E_{\rm inc}$ up to about $E_F$\ \cite{manduci}, 
in full agreement with the universal fusion excitation function 
of Refs.\ \cite{eudes}.
According to this systematics, for mass-symmetric systems the fusion 
extinguishes at $E_{\rm inc}\!\approx\!50$ MeV/nucleon\ \cite{eudes}.
This is in apparent opposition to the above experimental $R_E^{\rm exp}$ 
value, which is below 0.8 in this energy range; see data dots in 
Fig.\ \ref{iso}.
The discrepancy comes from the event selection essentially based on 
multiplicity cuts assuming a biunivocal correspondence between the 
reaction violence, i.e., the multiplicity, and the reaction centrality, 
i.e., the impact parameter $b$.
On one side, it has been demonstrated via model simulation that 
selecting events via multiplicity strongly mixes events of different 
impact parameters over a rather broad span in $b$\ 
\cite{zhang,bonnet,bas16}. 
That is the case even for the experimental subset of events which 
corresponds to the rather small fraction of the total reaction cross 
section $\sigma_R$.
Nominally, for the $R_E^{\rm exp}$ extraction used is the selection 
of the most central collision events, which corresponds to 
0.02\,$\sigma_R$\ \cite{lopez}.
On the other side, in this simulation the impact parameter together 
with $E_{\rm inc}$ completely determine the reaction mechanism.
A more quantitative comparison should include simulations with a wide 
impact parameter distribution and an event analysis closely simulating 
the experimental one, including the winnowing of simulation results 
by a filter of the experimental device.
This procedure, like the one performed in Ref.\ \cite{bas16}, is 
rather tedious and unnecessary here because of the already emphasized 
scope of the present paper.

One may argue that secondary emission due to the de-excitation of 
hot fragments may influence the $R_E$ value.
It is expected that fragments are formed in thermal equilibrium. 
Consequently, the secondary particles are emitted isotropically 
and, on the average, should not alter noticeably the value of $R_E$. 

The simulation results on the stopping observable $R_E$ are not 
conclusive about the characterization of a specific $\nu$ value.
The complex dependence of $R_E^{\rm exp}$ on $E_{\rm inc}$, 
especially its increase for $E_{\rm inc}\!>\!E_F$ for heavier 
systems\ \cite{lehaut,lopez} could not be explained in other 
simulation works either\ \cite{iqmd1,li18}.
Because of the subtle dependence of $R_E^{\rm exp}$ on details of 
event selection and particle choice\ \cite{lehaut,lopez} it is 
of interest to look for another observable sensitive to the stopping 
power and which is much less event-selection dependent.

\subsection{\label{sec3b}Linear momentum transfer}
\begin{figure}[tb]
\includegraphics[width=7.cm]{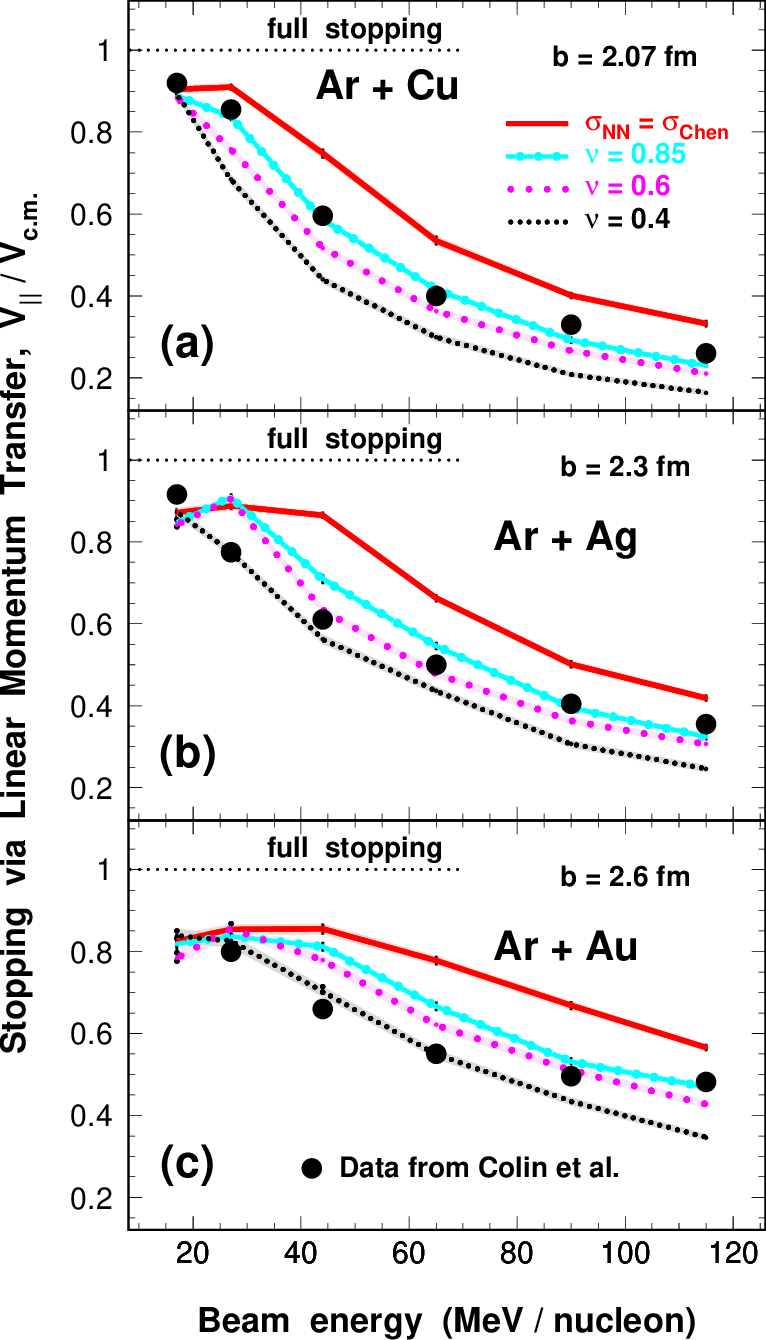}
\caption{(Color online.)
Linear momentum transfer for (a) $^{40}$Ar\,+\,$^{63}$Cu, (b) 
$^{40}$Ar\,+\,$^{107}$Ag, and (c) $^{40}$Ar\,+\,$^{197}$Au 
reactions as a function of $E_{\rm inc}$.
Lines represent simulation results with varied parameter $\nu$
of Eq.\ (\protect\ref{pnu}).
The value of impact parameter $b$ corresponds to $b_{\rm eff}$ 
for the assumed geometrical 0.08\,$\sigma_R$.
The dimmed background stands for the estimated errors.
Symbols represent experimental data\ \protect\cite{colin}.
\label{lmt}}
\end{figure}

The robust LMT observable is nothing but the velocity ratio of
the in-beam component of the targetlike fragment and center-of-mass
velocities $v^{\rm TL}_\parallel / V_{\rm c.m.}$.
Figure \ref{lmt} displays the LMT for the reactions 
$^{40}$Ar\,+\,$^{63}$Cu, $^{107}$Ag, and $^{197}$Au in the top, 
middle, and bottom panels, respectively, at $E_{\rm inc}$ between 
17 and 115 MeV/nucleon.
The heavy dots are from the experiment\ \cite{colin} and the various 
line types are the results of the simulation.
The representative impact parameter $b_{\rm eff}$ is obtained as 
$\sqrt{2}\,\,b_{\rm max}$\ \cite{bar19} assuming the equivalence of the 
geometrical 0.08\,$\sigma_R$, i.e.\ 8\,\% of the total reaction cross 
section, to the experimentally selected subset of the most central 
collisions in Ref.\ \cite{colin}.
Akin to the analysis of $R_E$\ \cite{lopez}, to evaluate the LMT in 
Ref.\ \cite{colin} the event selection is based on multiplicity cuts 
and, as in the other case, it suffers a broad impact parameter 
contribution much beyond the estimated $b_{\rm max}$ of 0.08\,$\sigma_R$.
From Fig.\ \ref{lmt} it appears that all predictions correctly follow 
the general trend of the data.
Simulation results endorse the need for the in-medium reduction 
of the free-space $NN$ cross section.
However, the degree of agreement between the data and the simulation
is both system-size and energy dependent.
Similarly to the results on the observable $R_E$ the simulation 
indicates higher fusion contribution at $E_{\rm inc}\!\lesssim\!E_F$ 
in comparison to what is observed in the subset of central events 
selected for the experimental analysis of LMT, an effect which 
becomes stronger as the target mass increases.
Consequently,
the $\nu\!=\!0.85$ curve closely follows experimental points in the 
case of Ar+Cu reaction at all $E_{\rm inc}$ whereas for the two 
heavier systems, and depending on $E_{\rm inc}$, calculations with 
the other $\nu$ values are also compatible with the data.

\begin{figure}[tb]
\includegraphics[width=7.cm]{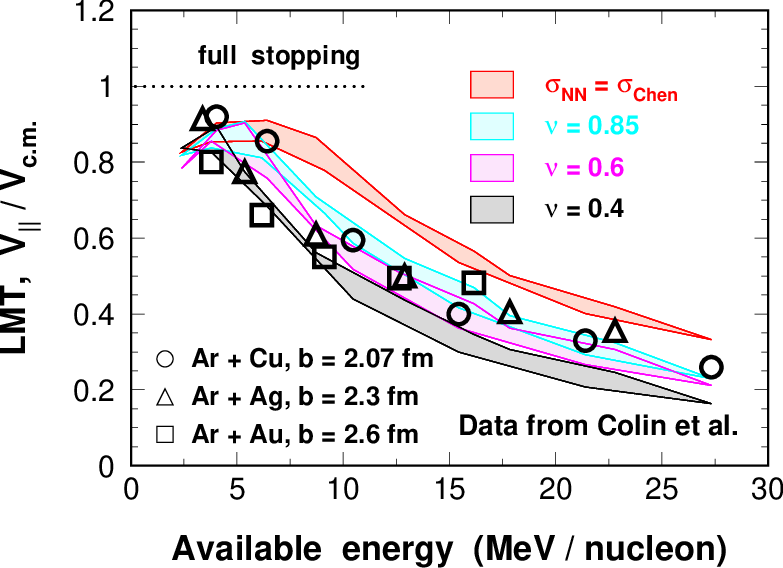}
\caption{(Color online.)
Same as Fig.\ \protect\ref{lmt} but as a function of $E_{\rm avail}$.
Each of the colored zones is due to the simulation with the given value 
of the parameter $\nu$ or for $\sigma_{\rm NN}\!=\!\sigma^{\rm free}$. 
The broken line delimiting a given zone corresponds to the simulation 
result including all three reactions studied.
\label{lmt_eav}}
\end{figure}

As was noticed above, the global cross-section reduction factor
${\cal F}$ of Eq.\ (\ref{fct_f}) deduced for the $\nu\!=\!0.85$ case, 
in contrast to the other two studied $\nu$ values, 
closely corresponds to the experimentally deduced ${\cal F}^{\rm exp}$; 
cf.\ in Fig.\ 10 of Ref.\ \cite{lopez} as well as in Fig.\ \ref{sig_f}.
One should keep in mind that factor ${\cal F}$ is connected with 
individual $NN$ diffusion processes, therefore with the c.m.\ energy 
of the interacting pair of nucleons or with the so-called available 
energy of a system, 
\begin{equation}
E_{\rm avail}=\frac{E_{\rm c.m.}}{A_{\rm sys}}=
E_{\rm inc}~\frac{A_p A_t}{(A_p + A_t)^2},
\label{eqeav}
\end{equation}
\noindent
where $A_p$, $A_t$, and $A_{\rm sys}$ are projectile, target and system 
mass, respectively.
In the case of mass-symmetric systems it scales to $E_{\rm inc}$ by a 
constant 4 so that the abscissa of Fig.\ \ref{sig_f} should be multiplied 
by 0.25.
In order to also consider mass-asymmetric systems and conclude on factor 
${\cal F}$ it is mandatory to express the system energy in terms of 
$E_{\rm avail}$ for the different systems.
To this end Fig.\ \ref{lmt_eav} shows LMT as a function of $E_{\rm avail}$.
From this figure it follows that for all systems studied the observable LMT 
is best reproduced by $\nu\!=\!0.85$ if $E_{\rm avail}\!\gtrsim\!10$ 
MeV/nucleon.
That is exactly the range of confidence in the extraction of 
${\cal F}^{\rm exp}$; cf.\ in Fig.\ 10 of Ref.\ \cite{lopez}.

\begin{figure}[tb]
\includegraphics[width=7.cm]{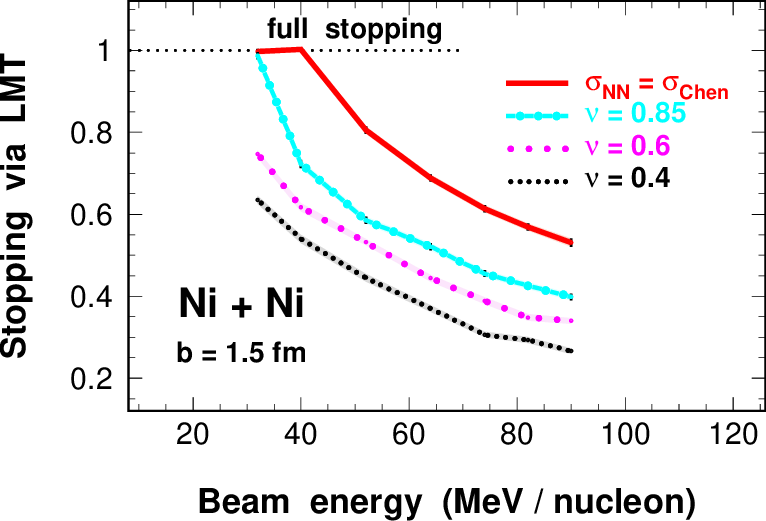}
\caption{(Color online.)
Linear momentum transfer for the $^{58}$Ni\,+\,$^{58}$Ni reaction 
at $b$\,=\,1.5 fm as a function of $E_{\rm inc}$.
Lines represent simulation results with varied parameter $\nu$
of Eq.\ (\protect\ref{pnu}).
The dimmed background stands for the estimated errors.
\label{lmt_1}}
\end{figure}

Experimental values on LMT for the two studied mass-symmetric reaction 
systems are not available.
However, in Fig.\ \ref{lmt_1} are displayed the simulation results on 
the LMT observable for the $^{58}$Ni\,+\,$^{58}$Ni reaction.
From the figure one infers that the overall behavior of LMT is not much 
affected by the mass symmetry.
In particular, at least for the nonreduced or moderately reduced 
$\sigma_{\rm NN}$ and at $E_{\rm inc}\!\lesssim\!E_F$, one observes the 
strong fusionlike behavior in which the fusion residue displaces at the 
c.m.\ velocity.
It would be of interest if the INDRA Collaboration could pay some attention 
to this observable in order to investigate its sensitivity to the details 
of the event selection.
Also, it would be challenging to explore how the two stopping 
observables are mutually interconnected.

We remind that both studied observables are sensitive to the choice 
of the impact parameter, a feature disregarded in the present 
qualitative presentation of the model.
In further calculations more elaborate confrontation of the model 
with various experimental observations is contemplated in which a 
true density dependence would be worth considering by introducing 
the local density in Eq.\ (\ref{pnu}).

\section{\label{sec4}Conclusions}
In this work we have revisited the bases of the extended-mean-field-like 
model DYWAN and presented current results in the framework of  
heavy-ion collisions in the range of incident energies up to around a 
hundred MeV per nucleon. 
Contrarily to the common usage of test-particles techniques in solving 
the 
Boltzmann-type transport problem, the specificity of our approach 
relies upon its intrinsic quantal background.
The static HF equation which serves to define the initial conditions 
of a colliding system, discussed in Sec.\ \ref{sec2} and mostly 
in Sec.\ \ref{sec2a}, is solved by expanding the system wave functions 
onto a set of spline wavelets which, in turn, are 
expressed as linear combinations of squeezed coherent states (CSs). 
This treatment provides a compact representation of SP wave 
functions and is illustrated in Sec.\ \ref{sec2a}.
Taking into consideration the characteristics of the corresponding 
ground states, i.e., the initial conditions for a one-body dynamical 
description of nuclei, using zero-range forces, the agreement with 
the respective experimental values is satisfactory.
The so-obtained wave functions are inputs into the dynamical 
TDHF-like equation as described in Sec.\ \ref{sec2b}.
The stability of the adopted numerical approach is proved by allowing a 
ground state nucleus to evolve dynamically for several hundreds of fm/$c$.
For an application to HIC at intermediate energies the one-body 
TDHF-like equation is complemented by a two-body dissipation kernel 
leading to an ETDHF-like equation as described in Sec.\ \ref{sec2c}.
This is performed via a so-called collision term dealing with
elastic scattering of couples of CS's which are subject 
to Pauli blocking and, in principle in-medium modified, 
elementary nucleon-nucleon ($NN$) cross section.
A procedure is proposed to retrieve the $N$-body information 
contained in the extended-mean-field description of HIC.
In other words, information on the nucleonic level may be extracted 
by organizing CSs in Slater determinants as described in
Sec.\ \ref{sec2d}.
This allows for encompassing the fragment formation 
and gives rise to genuine statistical fluctuations of physical 
observables, as discussed in the same subsection.

For the sake of illustration, in Sec.\ \ref{sec3} the model is 
applied to the two observables sensitive to the nuclear-stopping 
power: the energy isotropy ratio $R_E$ and the linear momentum 
transfer.
The need for the in-medium reduction of elementary $NN$ scattering 
is clearly confirmed.
Some discrepancy observed between simulation and experiment may be 
attributed, among other aspects, to the limited exploration of the 
simulation parameter space such as the role of impact parameter, 
choice of nuclear effective interaction, and accounting for the 
secondary emission of hot fragments.

The few existing intrinsically quantal approaches to the nuclear 
transport problem are all contingent on improvements.
In our case, at the level of the initial conditions, the paving of the 
phase space may be improved by achieving still better homogeneity 
of nuclear density. 
It might be obtained by a simple adjustment of the effective force 
used, but also by implementing other kinds of decomposition bases.
At the level of the dynamics, besides exploring the influence of the 
effective force chosen for the simulation, the collision term 
leaves a broad space for improvements.
Both the Pauli blocking and the treatment of in-medium 
modification of elementary NN collisions are questions fairly open 
to discussion and development.
A vast advancement may also come from further development of 
aggregation procedures.
An important upgrade of the model would be in the direction of 
inclusion of a stochastic force describing large density fluctuations.

\begin{acknowledgments}
Z.B.\ gratefully acknowledges the financial support and the
warm hospitality of the Facult\'e des Sciences of University
of Nantes and the Laboratory SUBATECH, UMR 6457.
The authors are indebted to Dr.\ B.\ Barker for valuable 
information on their work.
This work has been supported in part by French Science Foundation
CNRS-IN2P3 under Project \textit{Th\'eorie-ACME} and by Croatian
Science Foundation HrZZ under Project No. IP-2018-01-1257.
%% ACME (Au-delà du Champ Moyen Etendu) de l'IN2P3
\end{acknowledgments}
%-------------------------------------------------------------

\end{document}